\documentclass[prd,nofootinbib,aps,superscriptaddress,tightenlines,floats,floatfix]{revtex4}
\usepackage[]{youngtab}
\usepackage{epsfig}

\begin{document}

\title{Negative parity pentaquarks in large $N_c$ QCD and quark model}

\author{Dan Pirjol}
\affiliation{Center for Theoretical Physics, Massachusetts Institute of
Technology, Cambridge, MA 02139\vspace{4pt} }

\author{Carlos Schat}
\affiliation{CONICET and Department of Physics, Comisi\'on
Nacional de Energ\'{\i}a At\'omica, Avenida Libertador 8250,
(1429) Buenos Aires, Argentina\vspace{4pt} }

\date{\today}
\begin{abstract}
Recently, the $1/N_c$ expansion has been applied to the study of
exotic baryons containing both quarks and antiquarks. We extend
this approach to exotic states with mixed symmetric spin-flavor
symmetry, which correspond in the quark model to negative parity
pentaquarks, and discuss the large $N_c$ predictions for their
mass spectrum. The heavy exotics $\bar Q q^4$ transform as
$\mathbf{3}, \mathbf{\overline{6}},\mathbf{15}$ and $\mathbf{15'}$
under SU(3), while the light states $\bar q q^4$ include the
exotic  multiplets $\mathbf{\overline{10}},\mathbf{27,35}$. We
give mass relations among these multiplets in the $1/N_c$
expansion. In the quark model, the mass splittings between these
states are given by color-spin interactions. Using the observation
of an anticharmed exotic by the H1 collaboration, we give
predictions for the masses of other expected heavy pentaquarks.
\end{abstract}

\maketitle


\section{Introduction}

The $1/N_c$ expansion of QCD~\cite{'tHooft:1973jz,witten}
constrains the spin-flavor properties of baryons and their
couplings to mesons. In the $N_c \to \infty$ limit,  baryons have
an exact contracted spin-flavor symmetry
$SU(2F)_c$~\cite{Largenspinflavor,DJM}, which can be used to
classify states and organize the operator expansion in $1/N_c$.
This spin-flavor symmetry is broken at subleading orders in
$1/N_c$. The $1/N_c$ expansion has been applied to the
ground-state $[\mathbf{56},0^+]$ baryons
\cite{DJM,largenrefs,largen,heavybaryons}, and to their orbital
and radial
excitations~\cite{PY1,PY2,seventyminusrefs,Goity,SU3,radial,Cohen:2003jk,PiSc}.

Recently, this analysis was extended in Ref.~\cite{JM1,JM3,CoLe} to a new class of
exotic states, some of which are identified for $N_c=3$ with the pentaquark
states $q^4\bar q$ (the large $N_c$ expansion has been applied in \cite{hybrids}
to another class of exotic hadrons, the so-called `hybrid' states).
More generally, these states are labelled by their exoticness $E$, which
has a very simple interpretation in the quark model, wherein such an exotic
state contains $N_c+E$ quarks and $E$ antiquarks. Assuming a symmetric spin-flavor
wave function, the $E=1$ sector contains
the exotic representations $\mathbf{\overline{10}}, \mathbf{27}$
and $\mathbf{35}$ with positive parity.
The mass spectrum, axial couplings and strong widths of these states were
investigated in the $1/N_c$ expansion in~\cite{JM3}. They were alternatively
studied using the chiral soliton model \cite{Glozman,DPP,IKOR}, the diquark
model \cite{JW,KaLi,nuss}, the uncorrelated quark model
\cite{uncorr,Carlson:2003wc} and lattice QCD \cite{Thetaclat}.

Experimentally, several exotic candidates exist. They include the isosinglet
$\Theta^+(1540)$~\cite{Thetadiscovery} and two cascades with
$I=3/2$ $\Xi^{--}_{3/2}(1860)$ and
$I=1/2$ $\Xi^-(1855)$~\cite{xiexotics}. The  $\Theta^+(1540)$ is usually
assigned to the $\mathbf{\overline{10}}$ representation of SU(3).
Recently the H1 Collaboration
reported evidence for a narrow charmed state with a mass of about
3099 MeV \cite{H1}. The parity of these states is still undetermined, and
several methods have been proposed for their determination \cite{parity}.
However, later searches \cite{nocharm} could not confirm the H1 signal,
which leaves the existence of these states as an open issue for further experimental
work.

In this paper we extend the results of Ref.~\cite{JM3} by constructing a new
class of exotic states which correspond to different irreducible representations
of the contracted $SU(6)_c$ spin-flavor symmetry. The $E=1$ sector of these
new states contains the same exotic representations
$\mathbf{\overline{10}}, \mathbf{27}$ and $\mathbf{35}$, but with negative parity.
If the antiquark is heavy, the SU(3) representations are $\mathbf{3}$,
$\mathbf{\overline{6}}, \mathbf{15}$ and $\mathbf{15'}$.
We discuss the large $N_c$ predictions for the masses of these states.

The existence of these new states offers an alternative
interpretation of the observed pentaquark candidates, which can be
identified with the negative parity states. Although a negative
parity $\Theta^+(1540)$ would decay to $NK$ in an S-wave, which
would make the narrow width of this state difficult to explain,
such an interpretation can not be ruled out \cite{parity}. Another
attractive application of our results is to pentaquarks with a
heavy antiquark, for which negative parity states appear in a
natural way \cite{StWi}.

The paper is organized as follows. In Sec. II we construct the
negative parity exotic states for arbitrary $N_c$ and discuss the
qualitative predictions of the large $N_c$ limit following from
the $SU(4)_c$ contracted symmetry. In Sec. III the $1/N_c$
expansion is put on a quantitative basis, by deriving mass
operators for these states to subleading order in $1/N_c$ (for
heavy exotics) and to leading order in $1/N_c$ (for light
exotics). This is used to derive model-independent mass relations
among these states. In Sec. IV we present quantitative predictions
for the masses of the heavy states using the quark model with
spin-color interactions.  Finally, Sec.V discusses the
phenomenology of this model for the charmed pentaquarks with
negative parity, and presents predictions for other exotic states.

\section{Exotic states in the large $N_c$ limit}

We describe in this section the construction of the exotic states in the
large $N_c$ limit using the language of the constituent quark model.
The quark model is used simply as a bookkeeping device in order to enumerate
the states, and we do not make use of its usual dynamical assumptions.
Certain qualitative predictions of the large $N_c$ limit can be
obtained by counting the irreducible representations (towers) of the
contracted spin-isospin symmetry $SU(4)_c$ containing the states generated
from the
quark model construction. Such a tower is labelled by $K=0,1/2,1,\dots$ and
contains all states with spin and isospin satisfying $|I-J| \leq K \leq I+J$.
A more quantitative method for extracting the predictions
of the $1/N_c$ expansion is presented in the next section, in terms of quark
operators.
We stress that the existence of these states in the large $N_c$ limit is an
open question. However, assuming their existence, these methods give predictions
for their mass spectrum and couplings.

Baryons of exoticness $E$ contain $N_c+E$ quarks, and $E$ antiquarks. The wave
function of the $N_c+E$ ($E$) quarks (antiquarks) must be completely
antisymmetric. Consider for simplicity the case $E=1$. The color wave function of
the $N_c+1$ quarks must transform in a representation of
$SU(N_c)$ with two columns containing $N_c$ boxes and one box respectively,
since this is the only possibility which can give a color singlet after combining
it with the antiquark $\overline{\yng(1)}$. This implies that the
spin-flavor-orbital wavefunction of the $N_c+1$ quarks must transform
in the mixed symmetry representation of $SU(6)_q\times O(3)$. This constrains the
spin-flavor and orbital wave functions, which can transform in several ways,
corresponding to different representations of $SU(6)_q\times O(3)$

\begin{eqnarray}\label{spflorb}
\raisebox{-0.4cm}{\yng(5,1)}\quad  \to & &
\left[ \quad  \yng(6)\quad \otimes\quad
\raisebox{-0.4cm}{\yng(5,1)} \quad \right]\\
& & \oplus
\left[\quad  \raisebox{-0.4cm}{\yng(5,1)}
\quad \otimes\quad \yng(6)\quad \right] \quad
\oplus \quad \cdots\,.\nonumber
\end{eqnarray}

The first term corresponds to a completely symmetric SU(6)$_q$ spin-flavor wave function,
and describes the exotic baryons constructed in \cite{JM3}. Similar states are obtained in
the Skyrme model. The orbital wave function of these states must have one quark in an
excited state, with  the lowest states corresponding to a $p$-wave orbital excitation
$\ell=1$.  This means that the
system of $N_c+1$ quarks has negative parity, which yields positive parity exotics
after adding in the antiquark.

In this paper we focus on the states corresponding to the second term in
Eq.~(\ref{spflorb}). They have a completely symmetric orbital wave function,
with all $N_c+1$ quarks in an $s$-wave orbital. Note that there is only one
term in the sum on the right-hand side with this property. The absence of
orbital excitations in the orbital wave function means that these states are
expected to lie below\footnote{This is only true in the constituent quark model;
these states have no analogs in the
Skyrme model.} those obtained from the first term in Eq.~(\ref{spflorb}).
The orbital wave function of the $N_c+1$ quarks has positive parity, such that after
adding the antiquark, they correspond to negative parity exotics.
Such states were constructed in the uncorrelated quark model \cite{uncorr}
with light quarks (for reviews see \cite{review}), and in the diquark model
with one heavy antiquark \cite{StWi}.

We start by first reviewing the spin-flavor structure of the exotic states
constructed in \cite{JM3} which correspond to the first term in
Eq.~(\ref{spflorb}). This can be obtained from the $SU(6) \supset
SU(2)_{\rm spin} \times SU(3)_{\rm flavor}$ decomposition of the
completely symmetric state of the $q^{N_c+E}$ system.
In the $E=1$ sector this is given by
\begin{eqnarray}\label{SYM}
\yng(3)\cdots \yng(2) \quad \to & & [(J_q = 0) \otimes (0,\frac{N_c+1}{2}) ]
\oplus  [(J_q = 1) \otimes (2,\frac{N_c-1}{2})] \oplus
[(J_q = 2) \otimes (4,\frac{N_c-3}{2})] \oplus \cdots\\
\to & & [(J_q = 0) \otimes \mathbf{\overline{6}}]
\oplus [(J_q = 1) \otimes \mathbf{15}]
\oplus [(J_q = 2) \otimes \mathbf{15'}] \nonumber
\end{eqnarray}
with the last line corresponding to $N_c=3$. We denote the SU(3) representations
as usual by $(\lambda,\mu)$, with $\lambda$ the number of columns containing one
box, and $\mu$ the number of columns containing two boxes.
We show in  Fig.~1 the weight diagrams
of the SU(3) representations appearing in this decomposition.
Adding the antiquark and the orbital angular momentum
$L=1$ and keeping only representations
which can not annihilate into $\mathbf{8,10}$ gives the following
exotic states with $E=1$:
\begin{eqnarray}\label{exSYM}
\mathbf{\overline{10}}_{\frac12,\frac32}(\mathbf{\overline{6}_0})\,,\qquad
\mathbf{27}_{\frac12,\frac32,\frac12,\frac32,\frac52}(\mathbf{15}_1)\,,\qquad
\mathbf{35}_{\frac32,\frac52,\frac12,\frac32,\frac52,\frac72}(\mathbf{15}'_2)\,.
\end{eqnarray}
We show in brackets the spin-flavor quantum numbers of the $q^4$ system.
In the large $N_c$ limit, these states form two towers of degenerate
exotic baryons. This can be most easily seen by considering the states at the top
of the weight diagrams with quark content $\bar s q^{N_c+1}$.
Their spin is given by $\mathbf{J} = \mathbf{J}_q +
\mathbf{J}_{\bar q} + \mathbf{L} = \mathbf{I} + \mathbf{K}$ where
$\mathbf{K} = \mathbf{J}_{\bar q} + \mathbf{L} = 1/2, 3/2$.
This gives two towers with $K=1/2,3/2$, containing the $\bar s q^{N_c+1}$
members of the SU(3) representations

\begin{eqnarray}\label{towersS}
\begin{array}{c}
K=1/2 :\\
K=3/2 :\\
\end{array} & &
\begin{array}{c}
\mathbf{\overline{10}}_{\frac12}\,,\\
  \\
\end{array}
\begin{array}{c}
 \\
\mathbf{\overline{10}}_{\frac32}\,, \\
\end{array}\,\,
\begin{array}{c}
\mathbf{27}_{\frac12} \\
\mathbf{27}_{\frac12} \\
\end{array}
\,\,\,\,
\begin{array}{c}
\mathbf{27}_{\frac32}  \\
\mathbf{27}_{\frac32}  \\
\end{array}
\quad
\begin{array}{c}
 \\
\mathbf{27}_{\frac52} \\
\end{array}
\quad
\begin{array}{c}
 \\
 \mathbf{35}_\frac12 \\
\end{array}
\,\,\,
\begin{array}{c}
 \mathbf{35}_\frac32 \\
 \mathbf{35}_\frac32 \\
\end{array}
\quad
\begin{array}{c}
 \mathbf{35}_\frac52 \\
 \mathbf{35}_\frac52 \\
\end{array}
\,\,\,
\begin{array}{c}
  \\
 \mathbf{35}_\frac72 \\
\end{array}
\,,\cdots\\
\nonumber
(I,J) :& &(0,\frac12)\,, (0,\frac32)\,, (1,\frac12)\,,
 (1,\frac32)\,, (1,\frac52)\,, (2,\frac12)\,, (2,\frac32)\,,
(2,\frac52)\,, (2,\frac72)\,, \cdots
\end{eqnarray}
The multiplets contained in each tower are degenerate in the large $N_c$
limit. The first tower (with $K=1/2$) was
constructed in \cite{JM3} and contains all states degenerate with the
$\Theta^+(1540)$ (assumed to be in the spin 1/2 antidecuplet).

If the antiquark is heavy $\bar Q$, its spin decouples from the rest of the
hadron, and the spin of the light degrees of freedom $s_\ell$ becomes a good
quantum number. For each value
of $s_\ell \neq 0$ there is one heavy quark spin doublet with hadron spins
$J = s_\ell \pm 1/2$ (for $s_\ell \neq 0$), split by $1/m_Q$ effects.
The states considered here $\bar Q q^4$ have the spin of light degrees of freedom
$\mathbf{s}_\ell =
\mathbf{J}_q + \mathbf{L} = \mathbf{I} +\mathbf{K}$ with $K=1$.
This corresponds to a $K=1$ large $N_c$ tower, containing the SU(3)
representations
\begin{eqnarray}
K=1 & &:\mathbf{\overline{6}}_{1}\,,
\mathbf{15}_{0,1,2}\,, \mathbf{15'}_{1,2,3},
\end{eqnarray}
where the subscript denotes the spin of the light degrees of freedom
$s_\ell$ of the respective multiplet. This is different from the
heavy states considered in Ref.~\cite{JM3}, which were chosen to
belong to a $K=0$ tower.

\begin{figure}[hhh]
 \begin{center}
 \mbox{\epsfig{file=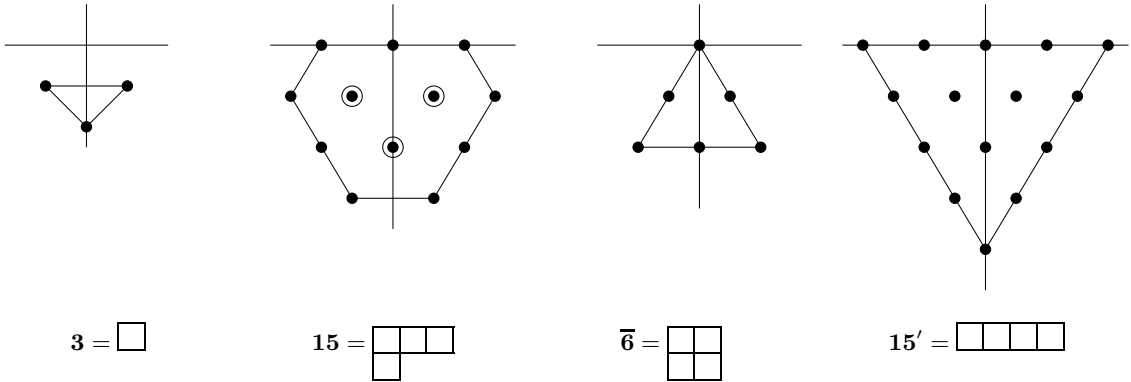,width=15cm}}\\[0.4cm]
$\mathbf{3}={\yng(1)}$ \hspace{2cm} $\mathbf{15}=\raisebox{-0.4cm}{\yng(3,1)}$
\hspace{2cm} $\mathbf{\overline{6}}=\raisebox{-0.4cm}{\yng(2,2)}$ \hspace{2cm}
$\mathbf{15'}={\yng(4)}$
 \end{center}
 \caption{
Weight diagrams for heavy pentaquarks with quark content $\bar Q q^4$.}
\label{fig2}
\end{figure}

We discuss next the spin-flavor structure of the states in the
mixed symmetric SU(6)$_q$ representation, corresponding to the second term
in Eq.~(\ref{spflorb}). The spin-flavor structure for this case
is considerably richer, and is very similar to that of the $L=1$ orbitally
excited baryons. The decomposition of the SU(6) spin-flavor wave function
into representations of $SU(2)_{\rm spin} \times SU(3)_{\rm flavor}$ can be
obtained e.g. by using the method presented in \cite{PY1}.
In the $E=1$ sector this decomposition contains the representations
(we assume here that $N_c$ is odd)
\begin{eqnarray}\label{MSspinflavor}
\raisebox{-0.4cm}{\yng(3,1)}\,
\cdots\, \yng(2)\quad  \to & & (J_q = 0) \otimes [(1,\frac{N_c-3}{2}) \oplus (2,\frac{N_c-1}{2})]\\
&+& (J_q = 1) \otimes [(0,\frac{N_c+1}{2}) \oplus (1,\frac{N_c-3}{2}) \oplus
(2,\frac{N_c-1}{2}) \oplus (3,\frac{N_c-5}{2}) \oplus (4,\frac{N_c-3}{2}) ]\nonumber\\
&+& (J_q = 2) \otimes [(2,\frac{N_c-1}{2}) \oplus (3,\frac{N_c-5}{2}) \oplus
(5,\frac{N_c-7}{2}) \oplus (6,\frac{N_c-5}{2}) ]\nonumber\\
&+& \cdots \nonumber
\end{eqnarray}
where the Young diagram on the left-hand side has $N_c+1$ boxes.
The ellipses correspond to unphysical representations for $N_c=3$.
Taking $N_c=3$ and keeping only the physical representations on the right-hand
side gives the possible states for the $q^4$ system
\begin{eqnarray}\label{q4}
\raisebox{-0.4cm}{\yng(3,1)}\quad  \to & & (J_q = 0) \otimes [ \mathbf{3} \oplus \mathbf{15}]
+ (J_q = 1) \otimes [\mathbf{\overline{6}} \oplus \mathbf{3} \oplus
\mathbf{15}' \oplus \mathbf{15} ]
+ (J_q = 2) \otimes [ \mathbf{15}  ]
\end{eqnarray}
The weight diagrams of these representations are shown in Fig.~\ref{fig2}.
The spin-flavor structure of these states is richer than that of the symmetric
spin-flavor states in Eq.~(\ref{SYM}).

In the diquark model \cite{JW}, the $q^4$ system is built from two $[qq]$
diquarks. Each of the diquarks transforms as $\mathbf{\overline{3}}$ under color,
and can be either `good' ($\mathbf{\overline{3}}_{S=0}$) or `bad' ($\mathbf{6}_{S=1}$)
according to their transformation under SU(3)$\times $SU(2) flavor-spin.
Because of their Bose statistics, the spin and flavor quantum numbers of systems
containing two identical diquarks are constrained in a specific way. However,
all states in Eq.~(\ref{q4}) are reproduced in the diquark model as well,
with diquark content as shown below
\begin{eqnarray}\label{2quark}
\mbox{good-good}: & & (J_q = 0) \otimes  \mathbf{3}\\
\mbox{good-bad}: & & (J_q = 1) \otimes  [\mathbf{3} \oplus \mathbf{15} ]\\
\mbox{bad-bad}: & & (J_q = 0) \otimes  \mathbf{15}\,, \quad
(J_q = 1) \otimes  [\mathbf{\overline{6}} \oplus \mathbf{15'} ]\,, \quad
(J_q = 2) \otimes \mathbf{15} \,.
\end{eqnarray}

We consider first the negative parity light pentaquarks, with
quark content $\bar q q^4$. These states were first constructed by
Strottman in Ref.~\cite{Strottman}, where their mass spectrum was
studied using a quark model with spin-color interactions following
the approach in Ref.~\cite{RLJ}. Recent work has been mainly
focused on the antidecuplet $\mathbf{\overline{10}}$ states, which
include the $\Theta(1540)$ pentaquark state \cite{CQM} (see
Ref.~\cite{review} for a recent review of the literature).

We review the construction of the complete set of states in the next section,
and restrict ourselves here to the exotic representations obtained from this
construction. Keeping only the states which can not
annihilate into ordinary $L=1$ negative parity states ($\mathbf{1,8,10}$) one finds the following
$E=1$ exotic states with negative parity
\begin{eqnarray}\label{E1}
\mathbf{\overline{10}}_{\frac12,\frac32}(\mathbf{\overline{6}}_1)\,,\qquad
\{ \mathbf{27}_{\frac12}(\mathbf{15}_0)\,, \mathbf{27}_{\frac12}(\mathbf{15}_1) \}\,,\qquad
\{ \mathbf{27}_{\frac32}(\mathbf{15}_1)\,, \mathbf{27}_{\frac32}(\mathbf{15}_2) \}\,,\qquad
\mathbf{27}_{\frac52}(\mathbf{15}_2)\,,\quad \mathbf{35}_{\frac12,\frac32}(\mathbf{15}'_1)\,.
\end{eqnarray}
Somewhat surprisingly, there are fewer light exotic states with negative parity
than with positive parity (compare with Eq.~(\ref{exSYM})). The reason for this is that, although
the $q^4$ system  has more states, most of them produce nonexotic states
($\mathbf{1,8,10}$) after adding in the light antiquark.

In the combined large $N_c$ and SU(3) limit, these states form again two towers of
degenerate exotic baryons.  This can be seen by examining the exotics with
strangeness $+1$ (quark content $\bar s q^{N_c+1}$)
which contain the following states (denoted by the SU(3) multiplets to which
they belong)
\begin{eqnarray}\label{towersMS}
\begin{array}{c}
K=1/2 :\\
K=3/2 :\\
\end{array} & &
\begin{array}{c}
\mathbf{\overline{10}}_{\frac12}\,\,\\
  \\
\end{array}
\begin{array}{c}
 \\
\mathbf{\overline{10}}_{\frac32}\,\, \\
\end{array}\,\,
\begin{array}{c}
\mathbf{27}_{\frac12} \\
\mathbf{27}_{\frac12} \\
\end{array}
\,\,\,\,
\begin{array}{c}
\mathbf{27}_{\frac32}  \\
\mathbf{27}_{\frac32}  \\
\end{array}
\quad
\begin{array}{c}
 \\
\mathbf{27}_{\frac52} \\
\end{array}
\quad
\begin{array}{c}
 \\
 \mathbf{35}_\frac12 \\
\end{array}
\,\,\,
\begin{array}{c}
 \star \\
 \mathbf{35}_\frac32 \\
\end{array}
\quad
\begin{array}{c}
 \star \\
 \star \\
\end{array}
\qquad
\begin{array}{c}
  \\
 \star \\
\end{array}
\,,\cdots\\
\nonumber
(I,J) :& &(0,\frac12)\,, (0,\frac32)\,, (1,\frac12)\,,
 (1,\frac32)\,, (1,\frac52)\,, (2,\frac12)\,, (2,\frac32)\,,
(2,\frac52)\,, (2,\frac72)\,, \cdots
\end{eqnarray}
This argument appears to indicate that the light pentaquarks with
negative parity form two sets of degenerate states in the large
$N_c$ limit, corresponding to the two irreducible representations
of $SU(4)_c$ with $K=1/2,3/2$. In the next section we will show
using the quark operator method, that there is an accidental
degeneracy between these two irreps, which is broken only by
$O(1/N_c)$ effects. This mass pattern is again different from that
obtained for the spin-flavor symmetric states in Ref.~\cite{JM3}
(see Eq.~(\ref{towersS})), for which the two towers are split by a
$O(1)$ mass difference.

In addition, there is another difference at $N_c=3$ coming from the fact that
in the real world the tower structure is not complete in the $I=2$ sector.
More precisely, four of the $I=2$ states (belonging to a $\mathbf{35}$)
expected to exist for $N_c\geq 5$ do not exist in the physical
world with $N_c=3$. The missing states are shown above as stars.

This situation is somewhat analogous to that
of the negative parity $\Delta_{1/2,3/2}$ states in the $\mathbf{70}$ of
SU(6). These states can not be unambiguously assigned to a $SU(4)_c$ tower for the
physical $N_c=3$ case \cite{PY2,PiSc}. Therefore, assuming that the $\Theta^+(1540)$
belongs to a negative parity $\mathbf{\overline{10}}_\frac12$ multiplet,
large $N_c$ QCD predicts only the existence of two $\mathbf{27}$ multiplets
degenerate with it, but not of $\mathbf{35}$ as in the positive parity case (see
Eq.~(\ref{towersS})).
This prediction might be difficult to test, since all these states can decay to $KN$
in $S-$wave and the $\mathbf{35}$ states will likely be too broad to resolve from the
continuum.

The negative parity exotics with one heavy antiquark $\bar Q q^{N_c+1}$ have a simpler
spin-flavor structure, which can be read off directly from that of the
$q^{N_c+1}$ system in Eq.~(\ref{q4}). They contain heavy quark spin doublets
with the spin of the light degrees of freedom $s_\ell = J_q$ and total
hadron spin $\mathbf{J}=\mathbf{s}_\ell \pm 1/2$. In the nonstrange
sector, the resulting states form one $K=1$ large $N_c$ tower containing
the following degenerate states:
an isosinglet with $s_\ell = 1$ (belonging to a $ \mathbf{\overline{6}}$),
three isovectors with $s_\ell = 0,1,2$ (belonging to a $ \mathbf{15}$)
and one isotensor with $s_\ell = 1$ (belonging to a $ \mathbf{15'}$)
\begin{eqnarray}\label{K1}
K=1& &: \mathbf{\overline{6}}_{J_q=1}\,,
\mathbf{15}_{J_q =0,1,2}\,, \mathbf{15'}_{J_q = 1}\,.
\end{eqnarray}
The states in the triplet $\mathbf{3}_{J_q=0,1}$ contain at least one
strange quark, and do not appear in this tower.

The heavy pentaquarks containing one strange quark $\bar Q s q^{N_c}$ contain
three towers of states: two towers with $K=1/2$ and one tower with $K=3/2$, as shown
below
\begin{eqnarray}\label{K1/2}
\begin{array}{c}
K=1/2 :\\
K'=1/2 :\\
K=3/2 :\\
\end{array} & &
\left(
\begin{array}{c}
\mathbf{3}_{0}  \\
\mathbf{15}_{0} \\
 \\
\end{array}
\right)\,
 \left(
\begin{array}{c}
\mathbf{3}_{1}\\
\mathbf{15}_{1} \\
\mathbf{\overline{6}}_{1} \\
\end{array}
\right)
\begin{array}{c}
 \\
 \\
\mathbf{15}_{2} \,  \\
\end{array}
\begin{array}{c}
 \\
 \\
\mathbf{15}_{0}  \\
\end{array}
\left(
\begin{array}{c}
\star \\
\mathbf{15}_{1} \\
\mathbf{15'}_{1} \\
\end{array}
\right)
\left(
\begin{array}{c}
 \star \\
 \star \\
\mathbf{15}_{2}  \\
\end{array}
\right)\,,\cdots\\
\nonumber (I,J_q) :& & \ \ \, (\frac12, 0)\,\,\,\,\, \ \ \
(\frac12, 1)\,\,\,\, (\frac12, 2)\,\,
 (\frac32, 0)\,\,\,\, (\frac32, 1)\,\,\,\,\, \ \ \, (\frac32, 2)\,, \cdots
\end{eqnarray}
To explain this equation, consider first the SU(3) symmetry limit.
Then the states along a horizontal line belong to the large $N_c$ K
towers shown and are therefore degenerate in the large $N_c$ limit.
When SU(3) is broken, the physical states are mixtures of the
SU(3) representations in each bracket. However, the eigenvalues
still belong to three large $N_c$ towers.
Some of these states are unphysical for $N_c=3$ and are absent
(represented by a star).

Inspection of Eq.~(\ref{K1/2}) gives the following general results:
\begin{itemize}

\item In the unbroken SU(3) limit, the two towers $K'=1/2$ and $K=3/2$ become
degenerate, since they both contain states in the $\mathbf{15}_{0}$.
This argument does not constrain the mass of the $K=1/2$ tower, which does not
contain any states in common SU(3) multiplets with the other two towers.
In particular, this means that in the large $N_c$ limit the
$\mathbf{3}$ states can have a mass different from that of the
other SU(3) representations in Fig.~1.

\item The two isospin multiplets with $(I,J_q)=(1/2,2)$ and $(3/2,0)$
(the $S=-1$ members of the $\mathbf{15}_{J_q=0}$ and $\mathbf{15}_{J_q=2}$)
are split only at $O(1/N_c)$ to all orders in SU(3) breaking.

\item The tower content constrains the pattern of SU(3) breaking in a very
specific way. For example, the three $(I,J_q)=(\frac12,1)$ states mix as an
effect of SU(3) breaking. However, the eigenvalues of the mass matrix are
constrained to coincide in the large $N_c$ limit with the masses of the
corresponding tower states.

\end{itemize}

To summarize this discussion, we showed that in the large $N_c$ limit the
heavy pentaquarks with negative parity fall into two groups of degenerate
states. The first group includes the five SU(3) multiplets containing the nonstrange
states ($\mathbf{\overline{6}_1,15_{0,1,2},15'_1}$), and the second group
contains the two $\mathbf{3}_{0,1}$ multiplets. This is different from the prediction of the
quark model with SU(6) symmetry, according to which all these states are
degenerate into a $\mathbf{210}$ of SU(6). The reason for this difference is
that in large $N_c$ QCD SU(6) is broken down to SU(6)$_c$ already at leading
order. The multiplets of SU(6) break down into irreducible representations of
the contracted symmetry (towers), which may or may not be degenerate at leading
order in $1/N_c$.

Although this phenomenon does not occur for ground state baryons,
where SU(6) spin-flavor symmetry is manifest in the mass spectrum in the
large $N_c$ limit,
a similar situation is encountered for the orbitally excited
$L=1$ baryons. For this case the $\mathbf{70}$ of SU(6) breaks
down in the large $N_c$ limit into three towers of nonstrange states plus
two additional towers containing the $\Lambda_{1/2,3/2}$ states
\cite{Goity,SU3,PY1,PY2,PiSc,Cohen:2003jk}.

In the next section we will formulate the large $N_c$ predictions in a more
quantitative way by writing down the mass operator of the negative parity states.

\section{Mass relations from the $1/N_c$ expansion}

The properties of exotics with mixed symmetry spin-flavor wave
functions can be studied quantitatively in the $1/N_c$ expansion
using the quark operators method \cite{DJM,largen}. The
spin-flavor algebra is realized in terms of operators acting on
the $ SU(2f) $  quark $\yng(1)$ and antiquark $\overline{\yng(1)}$
degrees of freedom. The hadron states are constructed by combining
quark and antiquark one-body states, with the proper quantum
numbers. Finally, the large $N_c$ expansion of any operator is
given by the most general combination of quark operators acting on
these states, to any given order in $1/N_c$.

We turn now to the construction of the exotic states with mixed
symmetry spin-flavor. This can be done in close analogy to that of
the orbitally excited baryons discussed in detail in
\cite{Goity,PY1,PY2}, so we will only give here the relevant
steps. The spin-flavor wave function of the $q^{N_c+1}$ states
transforms under $SU(2f)$ in the $MS_{N_c+1}$ representation
Eq.~(1). Under $SU(2f) \to SU(2)\times SU(f)$, this representation
contains all states with spin and isospin $S,I$ satisfying $|S-I|
\leq 1$ (except $S=I=0$ and $\frac12(N_c+1)$).

These states can be constructed explicitly as the combination of a
system of $N_c$ quarks with symmetric spin-flavor symmetry (core) with spin and
isospin $S_c=I_c$, plus an additional quark (excited)
\begin{eqnarray}\label{SI}
|SI\rangle = c_+ |S_c=I_c=I+\frac12\rangle \otimes |\frac12\rangle +
c_- |S_c=I_c=I-\frac12\rangle \otimes |\frac12\rangle
\end{eqnarray}
One distinguishes two situations: a) if $S\neq I$, only one term appears in
this sum, according to $S - I = 1$ ($c_+=1, c_-=0$) or $I-S=1$
($c_+=0, c_-=1$); b) if $S=I$ then both terms appear, and the coefficients
$c_\pm$ are determined by requiring that the state $|SI\rangle$ transforms
according to the $MS_{N_c+1}$ representation of $SU(4)$. One finds
for both SU(2) and SU(3) flavor symmetry \cite{Goity}
\begin{eqnarray}
c_+(I=S) = \sqrt{\frac{S}{2S+1}}\sqrt{\frac{N_c+2S+3}{N_c+1}}\,,\qquad
c_-(I=S) = -\sqrt{\frac{S+1}{2S+1}}\sqrt{\frac{N_c-2S+1}{N_c+1}}\,.
\end{eqnarray}
We took into account the fact that for this case the core contains
$N_c$ quarks, instead of $N_c-1$ as for the orbitally excited
baryons. Finally, the exotic state is obtained by combining the
state (\ref{SI}) with the antiquark $\bar q$ with appropriate
total quantum numbers. We will consider here the negative parity
exotics with one heavy antiquark, for which the nontrivial
spin-flavor structure is carried by the $q^{N_c+1}$ system alone
\begin{eqnarray}
|\Theta; JIS; m\alpha\rangle = \sum_{m_1 m_2} |SI;m_1\alpha\rangle  |\bar Q;m_2\rangle
\langle Jm| S\frac12; m_1 m_2\rangle
\end{eqnarray}
In the heavy quark limit, both the total spin $J$ and the spin of the light
degrees of freedom $S$ are good quantum numbers and the exotic states form
heavy quark spin doublets with $J= S\pm\frac12$ \cite{IsWi}.

Physical operators such as the Hamiltonian, axial currents, etc. can be
represented in the $1/N_c$ expansion by operators acting on the quark
basis states constructed as above. We will consider here in some detail the $1/N_c$
expansion of the mass operator ${\cal M}$. Keeping terms up to $O(1/N_c)$, this
has the general form
\begin{eqnarray}\label{op}
{\cal M} = N_c c_0 {\bf 1} + \sum_i {\cal O}_i^{(0)} + \sum_i
{\cal O}_i^{(1)} + \cdots\,,
\end{eqnarray}
where ${\cal O}^{(j)}$ are the most general isoscalar and Lorentz
scalar operators whose matrix elements scale like $1/N_c^j$.

The rules for constructing quark operators for states containing
both quarks and antiquarks have been formulated in
Ref.~\cite{JM3}. We will consider here only exotics containing one
antiquark (exoticness $E=1$). The building blocks for constructing
the most general operator are: a) antiquark operators
$\Lambda_{\bar q} : T_{\bar q}^a, S_{\bar q}^i, \bar G_{\bar
q}^{ia}$, acting on the antiquark degrees of  freedom and b)
operators $\Lambda_q$ acting on the states of the $N_c+1$ quarks
constructed above in Eq.~(\ref{SI}).

The quark building blocks include operators acting on both the
excited quark and on the core
\begin{itemize}

\item `excited quark' operators $s^i, t^a, g^{ia}$

\item `core' operators $S_c^i, T^a_c, G_c^{ia}$
\end{itemize}
We denote with $T^a$ the generators of the flavor group $SU(f)$, which is left
at this stage unspecified.

At leading order $O(N_c^0)$ in the $1/N_c$ expansion there is only one operator
contributing to the mass operator in Eq.~(\ref{op}), describing interactions between
the antiquark and the $N_c+1$ quarks
\begin{eqnarray}\label{Nc0}
O(N_c^0): \qquad O_0 = \frac{1}{N_c} T^a_{\bar q} S^i_{\bar q} G^{ia}
\end{eqnarray}
The only such operator containing only quark operators $\frac{1}{N_c} g^{ia} G_c^{ia}$
can be rewritten in terms of the unit operator and the $O(1/N_c)$ operators
$t^a T_c^a$ and $s^i S_c^i$ using the operator rules given in Ref.~\cite{Goity}.

At subleading order in $1/N_c$ there are more operators. A complete basis containing
only quark operators can be chosen as
\begin{eqnarray}
O(N_c^{-1}):&&  \qquad O_1 = \frac{1}{N_c} S_c^2\,,\quad
O_2 = \frac{1}{N_c} S^2\,,\quad O_3 = \frac{1}{N_c} T^2\,,\quad
O_4 = \frac{1}{N_c^2} t^a \{S_c^i\,, G_c^{ia}\}\,,\quad
O_5 = \frac{1}{N_c^2} g^{ia} T_c^a S_c^{i}
\end{eqnarray}
There are other  two-body operators which can be written,
such as $s^i S_c^i$ and $t^a T_c^a$. They can be
eliminated using the identities $S^2 = S_c^2+s^2 + 2s^i S_c^i$
and $T^a T^a = T_c^2 + t^2 + 2t^a T_c^a$.
The operator $T_c^2$ can be related to the core spin operator $S_c^2$.
For two light flavors they are equal $T_c^2 = S_c^2$, while for $f= 3$ they are
related as $T_c^2 = S_c^2 + N_c(N_c+6)/12$ (see Ref.~\cite{SU3}).

The three-body operators are linearly independent only for $f\geq 3$.
For two light flavors the operator $O_4$  can be reduced to two-body operators by
using the identities \cite{DJM,largen} (this assumes a core containing $N_c$ quarks)
\begin{eqnarray}\nonumber
2\{S_c^i\,, G_c^{ia}\} = (N_c+2) I_c^a\,,\qquad
\end{eqnarray}
which gives
\begin{eqnarray}
O_4 = \frac{1}{N_c^2} i^a \{S_c^i\,, G_c^{ia}\} =
\frac{N_c+2}{2N_c^2} i^a I_c^a = \frac{N_c+2}{4N_c^2}\left( I^2 -
S_c^2 - \frac34\right)
\end{eqnarray}
The three-body operator $O_5$ is subleading for $f=2$
\begin{eqnarray}
\frac{1}{N_c^2} g^{ia} T_c^a S_c^{ia} =
\frac{1}{4N_c^2}(S^2 - \frac34 - S_c^2)(I^2 - \frac34 - S_c^2) = O(1/N_c^2)
\end{eqnarray}

In the general case of $f\geq 3$ light flavors, $O_4$ can be rewritten
using an $SU(2f)$ operator identity for the core operators given in
Ref.~\cite{DJM} as
\begin{eqnarray}
t^a \{S_c^i\,, G_c^{ia} \} = \frac12 d^{abc} t^a \{ T_c^b\,, T_c^c\} -
\frac{1}{2f}(f-4)(N_c+f) t^a T_c^a\,.
\end{eqnarray}
In this form, the spin-independence of this operator becomes apparent:
although it formally depends on the core spin degrees of freedom, its
matrix elements depend in fact only on the flavor of the state (unless
such dependence is introduced indirectly through the spin-flavor
wave function Eq.~(\ref{SI})). Finally, the operator $O_5$ can be written
as
\begin{eqnarray}
O_5 = \frac{1}{4N_c^2}( S^2 - S_c^2 - \frac34)(T^2 - \frac{f^2-1}{2f} - T_c^2)
\to
\frac{1}{4N_c^2}( S^2 - S_c^2 - \frac34)[T^2 -  \frac{N_c^2+6N_c+16}{12}-S_c^2]
\quad (f=3)
\end{eqnarray}

One can  consider also SU(3) breaking effects. Keeping only $O(N_c^0)$
operators in the $1/N_c$ counting, there are three possible operators, given
by
\begin{eqnarray}
O(m_s): \qquad t^8\,,\qquad T_c^8\,,\qquad \frac{1}{N_c} d^{8ab} g^{ia} G_c^{ib}\,.
\end{eqnarray}
We will discuss in the following the predictions following from these
mass operators for negative parity exotics.

\subsection{Heavy pentaquarks mass relations}

We start by discussing first the simpler case of the
exotics with one heavy quark $\bar Q q^{N_c+1}$. In the heavy
quark limit the interactions of the heavy quark are suppressed by
$1/m_Q$, so that only quark operators need to be included in the Hamiltonian.
Working in the limit of SU(3) flavor symmetry and to subleading order
in $1/N_c$, the most general Hamiltonian reads
\begin{eqnarray}\label{Mop}
{\cal M} = N_c c_0 \mathbf{1} +  \frac{1}{N_c}\left[
c_1 S_c^2 + c_2 S^2 + c_3 T^2 + c_4  \frac{1}{N_c} t^a \{ S_c^i\,, G_c^{ia}\}
+ c_5 \frac{1}{N_c} g^{ia} S_c^i T_c^{a}
\right] + O(1/N_c^2)\,.
\end{eqnarray}
We list in Table I the matrix elements of these operators on the
heavy exotic states.

The qualitative structure of the mass spectrum can be understood
from this operator as follows. Although formally of $O(1/N_c)$,
the $SU(f)$ Casimir $ T^a T^a / N_c $ operator contains pieces of
$O(N_c), O(1)$ and $O(1/N_c)$ for $f > 2$. The $O(1)$ piece is
universal and can be absorbed into a redefinition of $c_0$, but
the $O(1)$ term takes different values on the nonstrange and the
$\mathbf{3}$ states. This introduces an $O(1)$ mass splitting
between these two types of states, as required by the tower
structure shown in Eq.~(\ref{K1/2}). The remaining operators have
matrix elements of $O(1/N_c)$.

\begin{table}
\begin{eqnarray}\nonumber
\begin{array}{c|cccccc}
\hline \mbox{state} &  S_c^2 & S^2 & I^2_{SU(2)} & T^2_{SU(3)} &
 t^a \{ S_c^i, G_c^{ia}\} &  g^{ia} S_c^i T_c^a  \\
\hline
\hline
 & & & & & & \\
\mathbf{3}_0 & \frac34 & 0   & - & \frac{1}{12}(N_c+1)^2
& -\frac14 & \frac18(N_c+6) \\
 & & & & & & \\
\mathbf{15}_0 & \frac34 & 0  & 2 & \frac{1}{12} (N_c^2+8N_c+31)
& \frac{N_c+3}{8} & -\frac{1}{16}(N_c+3)  \\
 & & & & & & \\
\hline \\
\mathbf{3}_1 & \frac34 & 2  &  - & \frac{1}{12}(N_c+1)^2
&  -\frac14  & -\frac{1}{24} (N_c+6)\\
 & & & & & & \\
\mathbf{\overline{6}}_1 & \frac34 & 2  & 0 &
\frac{1}{12}(N_c^2+8N_c+7)
&  -\frac{3N_c+5}{8}  & \frac{1}{48}(N_c-9)  \\
 & & & & & & \\
\mathbf{15}_1 & \frac{7N_c+23}{4(N_c+1)} & 2
& 2 & \frac{1}{12} (N_c^2+8N_c+31)
& -\frac{3N_c^2+26N_c+31}{24(N_c+1)} & -\frac{(N_c^2-2N_c-27)(N_c+9)}{48(N_c+1)^2}\\
 & & & &  & & \\
\mathbf{15'}_1 & \frac{15}{4} & 2   & 6 &
\frac{1}{12}(N_c^2+8N_c+79)
& \frac{3N_c+11}{8} & -\frac{5}{48}(N_c+9) \\
 & & & &  & & \\
\hline \\
\mathbf{15}_2 & \frac{15}{4} & 6   & 2 &
\frac{1}{12}(N_c^2+8N_c+31)
&  -\frac{5(N_c+1)}{8}  & \frac{1}{16}(N_c-15) \\
 & & & & & & \\
\hline
\end{array}
\end{eqnarray}
{\caption{ Matrix elements for the heavy exotic states.}}
\end{table}

The mass operator in Eq.~(\ref{Mop}) leads to mass relations
among the heavy exotics, valid up to $1/N_c^2$ corrections.
When restricted to the subspace of the nonstrange states (containing 5 states),
this operator contains 4 independent parameters. This gives one
model-independent mass relation
\begin{eqnarray}\label{mr1}
\mbox{(I)}\qquad \frac12 (\mathbf{15}_{2} - \mathbf{15'}_{1}) =
\mathbf{\overline{6}}_{1} - \mathbf{15}_{0} + O(1/N_c^2)\,.
\end{eqnarray}
One additional mass relation can be written
provided that the matrix elements of the
operator $S_c^2$ are evaluated at $N_c\to \infty$. This connects
the masses of the $ \mathbf{15}$ states as
\begin{eqnarray}\label{mr2}
\mbox{(II)}\qquad\frac23 \mathbf{15}_{0} + \frac13 \mathbf{15}_{2} =
\mathbf{15}_{1} + O(1/N_c^2)\,.
\end{eqnarray}
Both these relations assume only isospin symmetry.
A more complete discussion of the mass spectrum including SU(3) breaking
effects will be given elsewhere.

\subsection{Light negative parity pentaquarks}

We consider here the complete set of the $E=1$ light exotics with negative
parity, and study their mass spectrum at leading order in $1/N_c$.
The spin-flavor wave function of the $q^{N_c+1}$ system transforms in the
$MS_{N_c+1}$ representation of SU(6) and its decomposition into
representations of $SU(2)\times SU(3)$ spin-flavor has been given in
Eq.~(\ref{MSspinflavor}).

Adding in the antiquark transforming
as $\mathbf{\overline{6}} = (J_{\bar q} = \frac12) \otimes \mathbf{\overline{3}}$,
generates many states. They can be easily enumerated at $N_c=3$ using
the representation content of the $q^4$ system in Eq.~(\ref{q4}). We will divide
them into non-exotic $(\mathbf{1,8,10})$ and exotic states
$(\mathbf{\overline{10}, 27, 35})$. We use a notation which makes explicit the
spin-flavor transformation properties of the $q^{N_c+1}$ system:
$R_J(R^q_{J_q})$.

The nonexotic states are:
\begin{eqnarray}\label{nonex}
J=\frac12: & & \{ \mathbf{1}_\frac12 (\mathbf{3}_0)\,,
                  \mathbf{1}_\frac12 (\mathbf{3}_1) \}\\
\nonumber
& & \{ \mathbf{8}_\frac12 (\mathbf{3}_0)\,,
       \mathbf{8}_\frac12 (\mathbf{15}_0)\,,
       \mathbf{8}_\frac12 (\mathbf{\overline{6}}_1)\,,
       \mathbf{8}_\frac12 (\mathbf{3}_1)\,,
       \mathbf{8}_\frac12 (\mathbf{15}_1)\}\\
\nonumber
& & \{ \mathbf{10}_\frac12 (\mathbf{15}_0)\,,
       \mathbf{10}_\frac12 (\mathbf{15'}_1)\,,
       \mathbf{10}_\frac12 (\mathbf{15}_1)\}\\
\nonumber
J=\frac32: & & \mathbf{1}_\frac32 (\mathbf{3}_1)\,,\qquad
       \{ \mathbf{8}_\frac32 (\mathbf{\overline{6}}_1)\,,
       \mathbf{8}_\frac32 (\mathbf{3}_1)\,,
       \mathbf{8}_\frac32 (\mathbf{15}_1)\,,
       \mathbf{8}_\frac32 (\mathbf{15}_2)\}\\
\nonumber
& & \{ \mathbf{10}_\frac32 (\mathbf{15'}_1)\,,
       \mathbf{10}_\frac32 (\mathbf{15}_1)\,,
       \mathbf{10}_\frac32 (\mathbf{15}_2)\}\\
\nonumber
J=\frac52: & & \mathbf{8}_\frac52 (\mathbf{15}_2)\,,\qquad
\mathbf{10}_\frac52 (\mathbf{15}_2)\,.
\end{eqnarray}

The exotic states have been enumerated in Eq.~(\ref{E1}), and the
large $N_c$ predictions for their masses are described in terms of
the two tower structure in Eq.~(\ref{towersMS}). We include them
here again for completeness, making explicit also the quantum
numbers of the $q^4$ system for each state
\begin{eqnarray}\label{ex}
J=\frac12: & & \mathbf{\overline{10}}_\frac12 (\mathbf{\overline{6}}_1)\,,\qquad
               \{ \mathbf{27}_\frac12 (\mathbf{15}_0)\,,
                  \mathbf{27}_\frac12 (\mathbf{15}_1) \}\,,\qquad
\mathbf{35}_\frac12 (\mathbf{15'}_1)\\
\nonumber
J=\frac32: & & \mathbf{\overline{10}}_\frac32 (\mathbf{\overline{6}}_1)\,,\qquad
               \{ \mathbf{27}_\frac32 (\mathbf{15}_1)\,,
                  \mathbf{27}_\frac32 (\mathbf{15}_2) \}\,,\qquad
\mathbf{35}_\frac32 (\mathbf{15'}_1)\\
\nonumber J=\frac52: & & \mathbf{27}_\frac52 (\mathbf{15}_2)
\end{eqnarray}
In general, states with the same quantum numbers written within braces
$\{\cdots \}$ will mix.

The transformation properties of these states under the diagonal SU(6)
can be obtained by combining the representation of the
$q^{N_c+1}$ system with that of the antiquark. This produces three representations
\begin{eqnarray}\label{su6prod}
MS_{N_c+1} \otimes \mathbf{\overline{6}} = S_{N_c} \oplus MS_{N_c} \oplus
Ex_{N_c}
\end{eqnarray}
with $Ex_{N_c}$ the representation of SU(6) with the Young diagram
$[N_c+1,2,1,1,1]$ ($[n_1,n_2,n_3,n_4,n_5]$ denote the number of
boxes on each row). The nonexotic representations are the
symmetric representation $S_{N_c}$ with Young diagram
$[N_c,0,0,0,0,0]$, and the mixed-symmetric representation
$MS_{N_c}$ with Young diagram $[N_c-1,1,0,0,0,0]$. For $N_c=3$ the
exotic representation $Ex_3$ is the $\mathbf{1134}$. The
spin-flavor content of the first two SU(6) representations on the
right-hand side is well known
\begin{eqnarray}
S_{N_c} &=& \mathbf{8}_\frac12\,,\quad \mathbf{10}_\frac32\,,\cdots\\
MS_{N_c} &=& \mathbf{1}_\frac12\,,\quad
\mathbf{8}_\frac12\,,\quad \mathbf{8}_\frac32\,,\quad
\mathbf{10}_\frac12\,,\cdots\,.
\end{eqnarray}
The corresponding decomposition of $Ex_{N_c}$ can be obtained by
subtracting these states from the complete set in
Eqs.~(\ref{nonex}), (\ref{ex}). In particular, the exotic states
are all contained in $Ex_{N_c}$. For later reference, we give here
the values of the quadratic Casimirs for each of these SU(6)
representations
\begin{eqnarray}\label{Cas}
C_6(S_{N_c}) &=& \frac{5}{12}N_c(N_c+6)\\
C_6(MS_{N_c}) &=& \frac{1}{12}N_c(5N_c+18) \nonumber \\
C_6(Ex_{N_c}) &=& \frac{5}{12}(N_c^2 + 6 N_c + 12)\,.\nonumber
\end{eqnarray}

We would like to compute the mass spectrum of these states at leading order
in $N_c$. There are two operators which can appear in their mass operators
at $O(N_c^0)$.
They are the two-body $\bar q-q$ interaction introduced in
Eq.~(\ref{Nc0}), and  $\frac{1}{N_c}T^a T^a$, which
was seen to contain an enhanced $O(1)$ piece
\begin{eqnarray}
{\cal M} = N_c c_0 1 + c_1 \frac{1}{N_c} S_{\bar q}^i T^a_{\bar q} G_q^{ia}
+ c_2 \frac{1}{N_c} T^a T^a  + O(\frac{1}{N_c})\,.
\end{eqnarray}
The first operator is related to the generator of the diagonal SU(6) group
$G^{ia} = G_{\bar q}^{ia} + G_q^{ia}$ as
\begin{eqnarray}
2G_{\bar q}^{ia} G_q^{ia} = G^{ia} G^{ia} - G_q^{ia} G_q^{ia}
- G_{\bar q}^{ia} G_{\bar q}^{ia}\,.
\end{eqnarray}
The matrix element of $G^{ia} G^{ia}$ can be off-diagonal on the
space of the $q^{N_c+1}$ states, and in general it introduces
mixing among states with the same quantum numbers $R_J$ but
different $q^{N_c+1}$ quantum numbers. We will demonstrate this
explicitly below on the example of the $\mathbf{1}_\frac12
(\mathbf{3}_{0,1})$ states. This mixing is constrained by the
$U(6)_q\times U(6)_{\bar q}$ symmetry of QCD in sectors with both
quarks and antiquarks  in the large $N_c$ limit \cite{JM3}. Since
the correct spin-flavor symmetry of large $N_c$ QCD is the
contracted $SU(2F)_c$,  this symmetry is in fact $U(1)_{N_q}\times
SU(6)_{cq}\times U(1)_{N_{\bar q}}\times SU(6)_{c\bar q}$.

We showed at the end of Sec.~II that the SU(6) irrep $MS_{N_c+1}$ breaks
down into two irreps of the contracted symmetry $SU(6)_c$, containing
$\mathbf{\overline{6}_1,15_{0,1,2},15'_1}$ and $\mathbf{3}_{0,1}$, respectively.
This means for example, that the two sets of states in Eq.~(\ref{nonex})
$\{\mathbf{8}_\frac12 (\mathbf{3}_0)\,,\mathbf{8}_\frac12 (\mathbf{3}_1)\}$
and $\{       \mathbf{8}_\frac12 (\mathbf{15}_0)\,,
       \mathbf{8}_\frac12 (\mathbf{\overline{6}}_1)\,,
       \mathbf{8}_\frac12 (\mathbf{15}_1)\}$ do not mix in the large
$N_c$ limit.

In certain special cases such as those considered in Ref.~\cite{JM3}
the $O(1)$ operator  $\frac{1}{N_c}G_{\bar q}^{ia} G_q^{ia}$ can be reduced to
the $O(1/N_c)$ subleading operators $\frac{1}{N_c}J^2$ and $\frac{1}{N_c}J_q^2$.
This happens for states which can
be assigned to unique $SU(6) \supset SU(3) \times SU(2)$ and $SU(6)_q \supset
SU(3)_q\times SU(2)_q$ spin-flavor
representations for the $q^{N_c+1}$ and total system, respectively.
In particular, this is true for the exotic states
$R=\mathbf{\overline{10},35}$, which are obtained from $R_q =
\mathbf{\overline{6},15'}$ after adding in the antiquark, including
also the states constructed in \cite{JM3}. [It turns out to be true also
for the $\mathbf{27}_{1/2}$ and $\mathbf{27}_{3/2}$, which are admixtures of
$R^q_{J_q}=\mathbf{15}_{0}, \mathbf{15}_{1}, \mathbf{15}_{2}$.]
For these cases the matrix elements
of $\frac{1}{N_c}G_{\bar q}^{ia} G_q^{ia}$ can be computed in terms of the quadratic
Casimirs of SU(6) (denoted $C_6$) and SU(3) (denoted as $C_3$)
representations for the total and $q^{N_c+1}$ states
\begin{eqnarray}\label{general}
& &\langle R_J(R^q_{J'_q})|4 G_{\bar q}^{ia} G_q^{ia} |R_J(R^q_{J_q})\rangle =\\
& &\qquad \big[ C_6(R_J) - \frac13 J(J+1) - \frac12 C_3(R_J) \big] -
\big[ C_6(R^q_{J_q}) - \frac13 J_q(J_q+1) - \frac12 C_3(R^q_{J_q})\big] -2\nonumber
\end{eqnarray}
The negative parity exotic states in Eq.~(\ref{ex}) transform under SU(6) as
$R \sim Ex_{N_c}$, $R_q
\sim MS_{N_c+1}$ and under SU(3) as $R \sim \mathbf{\overline{10},27,35}$,
$R_q \sim \mathbf{\overline{6},15,15'}$. Using the results Eqs.~(\ref{Cas})
for the Casimirs of these representations for arbitrary $N_c$
we find the operator identity, valid on the space of the exotic states with
mixed symmetric spin-flavor wave function
\begin{eqnarray}\label{id1}
\left. 4 G_{\bar q}^{ia} G_q^{ia} \right|_{Ex(MS)} = \frac14 -
\frac13 \mathbf{J}^2 + \frac13 \mathbf{J_q}^2
\end{eqnarray}

For completeness, we give also the corresponding identity for the
symmetric spin-flavor states (considered in Ref.~\cite{JM3}). For
this case, the $q^{N_c+1}$ states transform under $SU(6)_q$ in the
$S_{N_c+1}$,  which after adding in the antiquark as an $ {\bf
\overline{6}}$ of $SU(6)$, gives the following representations of
the diagonal SU(6)
\begin{eqnarray}
S_{N_c+1} \otimes \mathbf{\overline{6}} = S_{N_c} \oplus Ex'_{N_c}
\ ,
\end{eqnarray}
where $Ex'_{N_c}$ corresponds to the Young diagram $[N_c +
2,1,1,1,1]$ and contains the exotic states. Its quadratic SU(6) Casimir
has the value $C_6(Ex'_{N_c}) = \frac{1}{12}(5N_c^2+42N_c+72)$.
Using Eq.~(\ref{general}) one finds for this case the same reduction
rule
\begin{eqnarray}\label{id2}
\left. 4 G_{\bar q}^{ia} G_q^{ia} \right|_{Ex'(S)} = \frac14 -
\frac13 \mathbf{J}^2 + \frac13 \mathbf{J_q}^2
\end{eqnarray}
We conclude that for both cases of mixed symmetry (\ref{id1}) and
symmetric (\ref{id2}) spin-flavor exotics,  the operator
$\frac{1}{N_c}G_{\bar q}^{ia} G_q^{ia}$ can be reduced to
$O(1/N_c)$ operators.

We consider next a case where such a simplification does not hold.
Consider the two nonexotic SU(3) singlet states with spin $J=\frac12$ in
Eq.~(\ref{nonex}). Under SU(6) they can transform as the $MS_{N_c}$ and
$Ex_{N_c}$,
with the two sets of states connected by a unitary transformation.
The quadratic Casimir of the diagonal spin-flavor SU(6) group
reads in the basis of these two states
\begin{eqnarray}
\sum_{A=1}^{35} T^A T^A = \frac13 \mathbf{J}^2 + \frac12 \sum_{a=1}^8 T^a T^a
+ 2 G^{ia} G^{ia} = \frac13 J(J+1) + \frac12 C_3(R)
+ 2G_q^{ia} G_q^{ia} + 2 + 4G_{\bar q}^{ia} G_q^{ia}
\end{eqnarray}
and is  diagonal in the basis $(\mathbf{1}_\frac12^{MS}\,,
\mathbf{1}_\frac12^{Ex})$, with eigenvalues given by the Casimirs of the respective
SU(6) representations.

Using these results, it is straightforward to compute the matrix
elements of the operator ${\cal O}_0 =
\frac{1}{N_c} G_{\bar q}^{ia} G_q^{ia}$ taken between the states
$(\mathbf{1}_\frac12(\mathbf{3}_0)\,, \mathbf{1}_\frac12(\mathbf{3}_1))$
(up to a two-fold ambiguity). The results are given in the Appendix.
Taking the large $N_c$ limit, the eigenvalues and eigenfunctions of the
mass operator on the subspace of the $\mathbf{1}_\frac12$ states are
\begin{eqnarray}\label{MMS}
\mathbf{1}_\frac12^{MS}(\mathbf{3}) &=& \mp\frac12 \mathbf{1}_\frac12(\mathbf{3}_0) +
\frac{\sqrt3}{2} \mathbf{1}_\frac12(\mathbf{3}_1)\,,\qquad
M_1 = N_c c_0 - \frac{3}{16} c_1\,,\\
\label{Mex}
\mathbf{1}_\frac12^{Ex}(\mathbf{3}) &=&
\pm \frac{\sqrt3}{2} \mathbf{1}_\frac12(\mathbf{3}_0) +
\frac12 \mathbf{1}_\frac12(\mathbf{3}_1)\,,\qquad
M_2 = N_c c_0 + \frac{1}{16} c_1\,.
\end{eqnarray}
Note that the large $N_c$ eigenstates do not have well defined
transformation properties under diagonal SU(6) for any finite
$N_c$ (although we labelled them with the corresponding irreps of
SU(6) into which they go for finite $N_c$).

A similar computation gives for the mass of the remaining SU(3) singlet
state
\begin{eqnarray}
\mathbf{1}_\frac32^{Ex}(\mathbf{3}_1): \qquad M = N_c c_0 + \frac{1}{16} c_1\,.
\end{eqnarray}
which turns out to be degenerate in the large $N_c$ limit with the
spin 1/2 exotic singlet Eq.~(\ref{Mex}).

These considerations are possibly of more than academic interest.
The singlet states $\mathbf{1}_{\frac12,\frac32}$ are expected
to be the lowest lying negative parity pentaquark states (assuming that such
states exist at all).
These states have the $q^4$ system in the $\mathbf{3}$ of flavor SU(3),
which is constructed in the diquark model from two good diquarks
(see Eq.~(\ref{2quark})). A similar result was also found in the quark model
computation of Strottman \cite{Strottman}, who estimated their masses to
lie around 1400 MeV. Two negative states are seen in this region
$\Lambda_\frac12(1406)$ and $\Lambda_\frac32(1519)$, both of which are
usually identified with the $L=1$ orbitally excited states. As shown
above, if the negative parity $q^4 \bar q$ states are stable,
three additional SU(3) singlet states are expected, two with spins $1/2$
and one with spin $3/2$. In the large $N_c$ limit, two of them are
degenerate in a pair with $J=1/2,3/2$.

We turn next to the light pentaquarks in exotic SU(3) representations.
Using the matrix elements in Appendix (or alternatively
the operator reduction rule Eq.~(\ref{id1})),
all negative parity exotic states in
the $\mathbf{\overline{10}}, \mathbf{27}$ and $\mathbf{35}$ turn out to be
degenerate at leading order in $1/N_c$, with a mass given by
\begin{eqnarray}
M_{Ex} = N_c c_0 + c_2 + O(1/N_c)\,.
\end{eqnarray}
These states are split only by $O(1/N_c)$ terms in the Hamiltonian which
were not considered here.
In particular, the two antidecuplet states $\mathbf{\overline{10}}_\frac12$
and $\mathbf{\overline{10}}_\frac32$ are predicted to be degenerate in the
large $N_c$ limit (and all states in the two towers $K=1/2,3/2$ in Eq.~(\ref{towersMS})
along with them). This follows from the absence of a $O(1)$ operator
which can distinguish between these states, and stands in contrast to the
situation in the symmetric spin-flavor states considered in Ref.~\cite{JM3}.
The latter states have nonzero orbital momentum, and a spin-orbit interaction term
$J_q^i \ell^i$ can introduce a $O(1)$ mass splitting between the
$J=1/2$ and $J=3/2$ antidecuplet states. We note however that $O(1/N_c)$
effects can easily produce a mass splitting of $\sim 200$ MeV, similar to the
$N-\Delta$ mass splitting, such that the two cases of positive and negative
parity could in fact have very similar mass spectra.

\section{Quark model predictions}

In the remainder of this paper we will study the negative parity exotic states in
some
detail using the constituent quark model with arbitrary number of colors $N_c$,
restricting ourselves to the $E=1$ states.
The color part of the wave function of the $N_c+1$ quarks must transform in the
fundamental representation and can be written as
\begin{eqnarray}
\chi_i^a = \frac{1}{\sqrt{N_c !}} \varepsilon_{a_1 a_2 \cdots (a_i) \cdots a_{N_c+1} }
|q_1^{a_1} q_2^{a_2} \cdots q_i^{a} \cdots q_{N_c+1}^{a_{N_c+1}} \rangle\,,
\qquad i = 1,\cdots, N_c
\end{eqnarray}
where the index in brackets $(a_i)$ is to be omitted. They are normalized as
\begin{eqnarray}
\langle \chi_j^b | \chi_i^a \rangle = \left\{
\begin{array}{ccc}
\delta_{ab}\,, & i=j &  \\
\frac{1}{N_c}p_{ij} \delta_{ab}\,, & i\neq j & (p_{ij}=(-)^{i-j-1})\\
\end{array}
\right.
\end{eqnarray}
The spin-flavor wave function of the $N_c+1$ quarks transforms
in the mixed symmetry representation. The corresponding spin-flavor
wave functions are identical to those for orbitally excited baryons, and
can be found in \cite{PY1} for the case of arbitrary $N_c$. They are constructed by
adding one quark $q_i$ to a symmetric state state of $N_c$ quarks with spin
and isospin $I_c=S_c$
\begin{eqnarray}
|S,I; m\alpha \rangle = \sum_{m_3,i_3}
|I_c,m_1\alpha_1\rangle \times |\frac12, m_2\alpha_2\rangle_i
\langle S m|I_c\frac12; m_1 m_2\rangle
\langle I\alpha|I_c\frac12; \alpha_1\alpha_2\rangle
\,.
\end{eqnarray}
Finally, the orbital wave function is completely symmetric and can be
written using a Hartree representation as the product of one-body wave functions
$\Phi(\vec r_1,\cdots, \vec r_{N_c+1}) = \phi_S(\vec r_1)
\cdots \phi_S(\vec r_{N_c+1})$. Putting together all factors, the complete
wave function of an $E=1$ exotic baryon with mixed symmetry spin-flavor structure
(negative parity) can be written as
\begin{eqnarray}\label{wf}
|\Theta_{\rm MS} ;JI,m\alpha \rangle = \frac{1}{\sqrt{N_c(N_c+1)}} \sum_{i=1}^{N_c+1}
\chi_i^a |\bar Q^a, m_3\rangle |SI;m-m_3,\alpha\rangle_i
\langle Jm|S\frac12;m-m_3,m_3 \rangle
\Phi(\vec r_1,\cdots ,\vec r_{N_c+1})
\end{eqnarray}
For simplicity, we took the antiquark to be an infinitely heavy quark $\bar Q$,
which does not introduce an additional orbital motion.
A similar construction gives also the wave function of an exotic with
strangeness $+1$ $q^{N_c}\bar s$.

The relevant Hamiltonian for the heavy exotic states $\Theta_{\bar
Q}$ describes the interactions of the nonrelativistic quarks with
the gluon field, plus the pure glue term
\begin{eqnarray}
{\cal H} = {\cal H}_{\rm kin} + {\cal H}_{\rm glue} + {\cal H}_{\rm q-q}\,.
\end{eqnarray}
The Coulomb quark-quark interaction, together with the kinetic
term $\sum_n {\cal H}_{\rm kin}^n = \sum_n \frac{1}{2m_n}(\vec p - g\vec A(x_n))^2$
and the pure glue term, give the dominant contributions in the large $N_c$ limit
\begin{eqnarray}\label{H}
{\cal H}_{\rm q-q} =
\frac{g^2}{4\pi}\sum_{m< n} \frac{T_m^x T_n^x}{|\vec r_n - \vec r_m |}
+ \frac{g^2}{4\pi}\sum_{n} \frac{T_n^x \bar T^x}{|\vec r_n |}\,.
\end{eqnarray}

The Hamiltonian ${\cal H}_{q-q}$ in Eq.~(\ref{H}) is  spin-flavor blind,
such that the heavy exotics with mixed-symmetric spin flavor $\Theta_{MS}$
constructed above fall into irreducible representations of the SU(6) group.
The lowest lying states are in the $\mathbf{210}$ of SU(6), which contains
all the representations of SU(3)$\times$SU(2) shown in Eq.~(\ref{q4}).
The degeneracy of these states is broken in the presence of the color
spin-spin hyperfine interaction \cite{CQM}
\begin{eqnarray}\label{Hhyp}
{\cal H}_{\rm hyp} = -V \sum_{i< j} (\lambda_i^a \lambda_j^a)(\vec \sigma_i
\cdot \vec \sigma_j)\,.
\end{eqnarray}

The eigenvalues of the hyperfine interaction can be computed using
SU(6) spin-color methods as explained in Ref.~\cite{RLJ}. The basic idea
is to decompose the color-spin-flavor wave function of the $q^4$ system into
irreducible representations of $SU(6)_{\rm sc}\times SU(3)_{\rm fl}$.
The generators of the color-spin group $SU(6)_{\rm sc}$ can be written in a
quark basis as
\begin{eqnarray}
S^i = \frac{1}{\sqrt3} \sum_{n=1}^{N_c+1} \frac{\sigma_n^i}{2}\,,\qquad
T^a = \frac{1}{\sqrt2} \sum_{n=1}^{N_c+1} \frac{\lambda_n^a}{2}\,,\qquad
F^{ia} = \sqrt2 \sum_{n=1}^{N_c+1} \frac{\sigma_n^i}{2}\cdot
\frac{\lambda_n^a}{2}\,.
\end{eqnarray}
With this definition, the generators are normalized as $\mbox{Tr}(\Lambda^A
\Lambda^B) = \frac12 \delta^{AB}$.
The generator $F^{ia}$ is simply related to the hyperfine Hamiltonian
Eq.~(\ref{Hhyp}), which is therefore diagonalized in terms of the quadratic
Casimir of the $SU(6)_{\rm sc}$
\begin{eqnarray}\label{C2def}
C_2(R_{\rm sc}) = \frac13 \mathbf{S}^2 + \frac12 T^a T^a + 2 F^{ia} F^{ia} =
\frac13 S(S+1) + \frac12 C_2(R_{\rm c}) + \frac32 C_F (N_c+1) +
\frac14 \sum_{i< j} (\lambda_i^a \lambda_j^a)(\vec \sigma_i
\cdot \vec \sigma_j)\,.
\end{eqnarray}

The exotics with symmetric orbital wave functions considered here
have a completely antisymmetric total color-spin-flavor wave
function of the $q^4$ system, which transforms as the
$\mathbf{3060}$ (for $f=3$) or $\mathbf{495}$ (for $f=2$) of
SU(6f). Decomposing it  into representations of $SU(6)_{\rm
sc}\times SU(3)_{\rm fl}$ and keeping only those $SU(6)_{\rm sc}$
representations which contain a $\mathbf{3}$ of color gives the
spin-color representations shown in Table II, corresponding to
each allowed SU(3) flavor representation. Using the relation
Eq.~(\ref{C2def}) for the quadratic Casimir of $SU(6)_{\rm sc}$
one finds the eigenvalues of the color-spin hyperfine interaction
\cite{RLJ}
\begin{eqnarray}\label{Hhypeigenvalue}
{\cal H}_{\rm hyp} = V \left[
\frac43 J(J+1) + 6(N_c + 1) C_F - 4 C_2(R_{\rm sc})\right]\,.
\end{eqnarray}
Here $C_2(R_{\rm sc})$ is the quadratic Casimir of the SU(6)
spin-color representation $R_{\rm sc}$ corresponding to a given
representation of flavor SU(3). The corresponding representations
and their Casimirs are given in Table II.

\begin{table}
\begin{eqnarray}\nonumber
\begin{array}{|c|c|c|}
\hline
R_{\rm fl} & R_{\rm sc} & C_2(R_{\rm sc}) \\
\hline
\hline
 & & \\
\mathbf{15'} &
\raisebox{-0.6cm}{\yng(1,1,1,1)} = \mathbf{15} \supset J_q=1 & \frac{14}{3} \\
 & & \\
\mathbf{15} &
\raisebox{-0.4cm}{\yng(2,1,1)} = \mathbf{105}\supset J_q=0,1,2 &
\frac{26}{3} \\
 & & \\
\mathbf{\overline{6}} &
\raisebox{-0.3cm}{\yng(2,2)} = \mathbf{105}\supset J_q=1 & \frac{32}{3} \\
 & & \\
\mathbf{3} & \raisebox{-0.2cm}{\yng(3,1)} = \mathbf{210}
\supset J_q=0,1 & \frac{38}{3} \\
 & & \\
\hline
\end{array}
\end{eqnarray}
{\caption{The representation of $SU(6)$ spin-color $R_{\rm sc}$
corresponding to each representation $R_{\rm fl}$ of $SU(3)$
flavor in Eq.~(\ref{q4}), and its quadratic Casimir $C_2(R_{\rm
sc})$.  For each representation of
 $SU(6)$ spin-color we show also the values of the quark spin
$J_q$ allowed under the decomposition $R_{\rm sc} \to
R_{SU(2)_{\rm sp}} \times  R_{SU(3)_c}$ (corresponding to a color
triplet $R_{\rm col} = \mathbf{3}$). }}
\end{table}

The mass formula predicts that the lowest eigenvalue is the
$\mathbf{3}_{J_q=0}$, which has the largest value for the spin-color SU(6)
Casimir. This result is in agreement
with the diquark model \cite{StWi}, according to which this state is composed
of 2 `good' scalar diquarks $\phi^{iA}$ (with $i$ a SU(3) flavor index and $A$
a SU(3) color index) in a relative $S-$wave
\begin{eqnarray}
|\mathbf{3}^i_{J_q=0}\rangle = \frac{1}{\sqrt{6!}}\varepsilon_{ABC}
\varepsilon_{ijk}
\bar Q^A \phi^{jB} \phi^{kC}
\end{eqnarray}
The remaining heavy exotics with negative parity lie above the triplet with
spin $J_q=0$, and their
mass splittings are given by one free parameter, the coupling $V$ in
Eq.~(\ref{Hhyp})
\begin{eqnarray}\label{hyp1}
& &\mathbf{3}_{J_q=1} - \mathbf{3}_{J_q=0} = \frac83 V\qquad (51\pm 3\mbox{
MeV})\\
& &\mathbf{\overline{6}}_{J_q=1} - \mathbf{3}_{J_q=0} = \frac{32}{3} V
\quad\,\, (203 \pm 11\mbox{ MeV})\\
& &\mathbf{15}_{J_q=0} - \mathbf{3}_{J_q=0} = 16 V\quad (304\pm 16\mbox{
MeV})\\
& &\mathbf{15}_{J_q=1} - \mathbf{3}_{J_q=0} = \frac{56}{3} V
\quad (355 \pm 19\mbox{ MeV})\\
& &\mathbf{15}_{J_q=2} - \mathbf{3}_{J_q=0} = 24 V\quad\,\, (456 \pm 24\mbox{
MeV})\\
\label{hyp6}
& &\mathbf{15'}_{J_q=1} - \mathbf{3}_{J_q=0} = \frac{104}{3} V
\quad (659\pm 35\mbox{ MeV})\,.
\end{eqnarray}
These results agree with Ref.~\cite{LeSi}.
The strength of the spin-spin coupling $V$ can be extracted from the $N-\Delta$ mass
splitting
\begin{eqnarray}
V = \frac{1}{16}(\Delta - N) = 18.3 \mbox{ MeV}
\end{eqnarray}
An alternative determination of $V$ from the $\Sigma_c -
\Lambda_c$ mass splitting gives a somewhat larger value
\begin{eqnarray}
V = \frac{3}{32}\left[ \frac13(\Sigma_c+2\Sigma_c^*) - \Lambda_c\right] =
19.8\mbox{ MeV}\,.
\end{eqnarray}
The numerical values for the mass splittings shown
in parantheses in Eqs.~(\ref{hyp1})-(\ref{hyp6}) used $V=19\pm 1$
MeV, which covers both these determinations.

These mass splittings are formally of $O(\alpha_s) \sim O(1/N_c)$. They agree with
the model independent predictions from the $1/N_c$ expansion Eq.~(\ref{Mop})
with a special relation among the $O(1/N_c)$ coefficients
\begin{eqnarray}
c_1 = 0\,,\qquad c_2 = \frac43 V\,,\qquad c_3 = 4 V\,,\qquad
c_{4,5} = 0\,.
\end{eqnarray}
The corresponding relation for two light flavors can be obtained
by noting that the Casimir $C_2(R_{\rm sc})$ in  Table II can be
written as $C_2(R_{\rm sc}) = 32/3 - I(I+1)$ for the nonstrange
states in the $\mathbf{\overline{6}},\mathbf{15},\mathbf{15'}$.
This summarizes the quark model mass formula on the subspace of
the nonstrange states as a two-parameter relation
\begin{eqnarray}
M(J_q,I) = \alpha + V \left[ \frac43 J_q(J_q+1) + 4 I(I+1) \right]
\end{eqnarray}
We will use in the next section experimental information on one of these states
to extract $\alpha$ and give predictions for all other states.

\section{Phenomenology}

The H1 Collaboration reported recently the observation of a narrow resonance in
the $D^{*-} p$ and $D^{*+} \bar p$ channels with mass and width
\begin{eqnarray}\label{H1}
M = 3099 \pm 3\mbox{ (stat)} \pm 5\mbox{ (stat) MeV}\,,\qquad
\Gamma = 12 \pm 3\mbox{ (stat) MeV}\,.
\end{eqnarray}
This resonance has been identified with an exotic state with quark
content $udud\bar c$. Neither the spin or the parity of this state have
been measured.
The phenomenology of this state has been studied using the large $N_c$ expansion
in Ref.~\cite{JM3} assuming that its parity is positive.

We will assume in this section that the state observed by the H1 Collaboration
has negative parity and explore the phenomenological implications of this
assumption following from the results of this paper.

We examine first predictions following solely from heavy quark symmetry.
The lowest lying pentaquarks with one charm antiquark $\bar c q^4$ with negative
parity have mixed symmetry spin-flavor wave functions and contain the
SU(3) representations shown in Eq.~(\ref{q4}). We will denote the nonstrange
states as $\Theta_{\bar c s_\ell}^{(R)}(J)$ with $s_\ell$ the spin of the light
degrees of the freedom, $J=s_\ell \pm 1/2$ is the total spin and $R$ is the SU(3)
representation of the state.

With this notation, the nonstrange states include
\begin{eqnarray}\label{I05q}
I = 0 &:& \Theta_{\bar c 1}^{(\mathbf{\overline{6}})}(\frac12,\frac32)\\
I = 1 &:& \Theta_{\bar c 0}^{(\mathbf{15})}(\frac12)\,,\quad
\Theta_{\bar c 1}^{(\mathbf{15})}(\frac12,\frac32)\,,\quad
\Theta_{\bar c 2}^{(\mathbf{15})}(\frac32,\frac52)\\
\label{I25q}
I = 2 &:& \Theta_{\bar c 1}^{(\mathbf{15'})}(\frac12,\frac32)\,.
\end{eqnarray}
In addition, there are also two SU(3) triplets with $s_\ell = 0$
and 1, containing at least one strange quark. They each contain
one isodoublet (with quark content $\{\bar c usud, \bar c udds\}$)
\begin{eqnarray}
T_{\bar c 0}(\frac12)\,,\qquad T_{\bar c 1}(\frac12,\frac32)
\end{eqnarray}
and one isosinglet (with quark content $\bar c usds$).

Assuming that these states can decay strongly into $D^{(*)} N$ channels,
heavy quark symmetry predicts ratios among the individual widths. These
predictions can be read off from Ref.~\cite{PY0}, where they were computed for the
corresponding strong decays of orbitally excited charmed baryons with negative parity.
For states above the $[D^-p]$ (2808 MeV) and $[D^{*-}p]$ (2948 MeV) thresholds (such
as the resonance at 3099 MeV observed by H1),
the decay ratios for the $S-$wave partial rates are (we denote here $\{\cdot \} =
\Gamma/|\vec p|$ the decay rate with the phase space factor removed)
\begin{eqnarray}\label{Swave}
& &\{\Theta_{\bar c 0}^{(\mathbf{15})}(\frac12) \to [N\bar D]_S\} \, : \,
\{\Theta_{\bar c 0}^{(\mathbf{15})}(\frac12) \to [N\bar D^*]_S\}  = 1\, :\, 3\\
& &
\{\Theta_{\bar c 1}^{(\mathbf{\overline{6},15})}(\frac12) \to [N\bar D]_S\}  \, : \,
\{\Theta_{\bar c 1}^{(\mathbf{\overline{6},15})}(\frac12) \to [N\bar D^*]_S\}  \, : \,
\{\Theta_{\bar c 1}^{(\mathbf{\overline{6},15})}(\frac32) \to [N\bar D]_S\}  \, : \,
\{\Theta_{\bar c 1}^{(\mathbf{\overline{6},15})}(\frac32) \to [N\bar D^*]_S\}  \\
& &\qquad = \frac34\, :\, \frac14\, :\, 0 \, :\, 1\,.\nonumber
\end{eqnarray}
Heavy quark symmetry predicts also the suppression of the $S-$wave amplitude
in the decay of the $s_\ell=2$ state
\begin{eqnarray}\label{suppress}
\Gamma\left(
\Theta_{\bar c 2}^{(\mathbf{15})}(\frac32) \to [N\bar D^*]_S\right) \sim
O(\Lambda^2/m_c^2)\,.
\end{eqnarray}
The corresponding predictions for the $D-$wave amplitudes can be found in
Ref.~\cite{PY0}, and we do not reproduce them here.
Some of the exotic states in Eq.~(\ref{I05q})-(\ref{I25q}) might lie above the
$\Delta \bar D$ (3101 MeV) and $\Delta \bar D^*$ (3242 MeV) thresholds.
These channels are also important for the decays of the $\mathbf{15}'$,
which can not decay to $N\bar D^{(*)}$ by isospin symmetry.
We give therefore
also the corresponding heavy quark symmetry predictions for the ratios of $S-$wave
partial widths into $\Delta \bar D^{(*)}$ channels
\begin{eqnarray}\label{Swave2}
& &
\{\Theta_{\bar c 1}^{(\mathbf{\overline{6},15})}(\frac12) \to [\Delta\bar D]_S\}  \, : \,
\{\Theta_{\bar c 1}^{(\mathbf{\overline{6},15})}(\frac12) \to [\Delta\bar D^*]_S\}  \, : \,
\{\Theta_{\bar c 1}^{(\mathbf{\overline{6},15})}(\frac32) \to [\Delta\bar D]_S\}  \, : \,
\{\Theta_{\bar c 1}^{(\mathbf{\overline{6},15})}(\frac32) \to [\Delta\bar D^*]_S\}  \\
& &\qquad = 0\, :\, 1\, :\, \frac38 \, :\, \frac58\nonumber\\
& &
\{\Theta_{\bar c 2}^{(\mathbf{15})}(\frac32) \to [\Delta\bar D]_S\}  \, : \,
\{\Theta_{\bar c 2}^{(\mathbf{15})}(\frac32) \to [\Delta\bar D^*]_S\}  \, : \,
\{\Theta_{\bar c 2}^{(\mathbf{15})}(\frac52) \to [\Delta\bar D]_S\}  \, : \,
\{\Theta_{\bar c 2}^{(\mathbf{15})}(\frac52) \to [\Delta\bar D^*]_S\}  \\
& &\qquad = \frac58\, :\, \frac38\, :\, 0 \, :\, 1\nonumber\,.
\end{eqnarray}

The resonance observed by H1 is seen in the $D^* p$ channel.
Assuming that future experiments see no signal in the $D N$
channel, the heavy quark predictions in Eq.~(\ref{Swave}) would
suggest identifying this resonance with a $J=\frac32$ state in the
$\mathbf{\overline{6},15}$. Since no other nearby states are
observed, the simplest assignment is the isosinglet $\Theta_{\bar
c 1}^{(\mathbf{\overline{6}})}(\frac32)$, decaying with equal
widths to $\bar D^{*-} p$  and  $\bar D^{*0} n$  (scenario 1).

An attractive possibility which naturally explains the small width
of the state (\ref{H1}) is to identify it with the $\Theta_{\bar c
2}^{(\mathbf{15})}(\frac32)$, whose S-wave decay width is
suppressed by heavy quark symmetry (see Eq.~(\ref{suppress})).
This state can only decay to $D^{(*)} N$ in $D-$wave. This is the
$I_3=0$ member of an isotriplet, and one expects to see two
similar nearby states, decaying to $\bar D^{*0} p$ (for the
$I_3=+1$ state) and to $D^{*-} n$ (for the $I_3=-1$ state). We
will refer to this as to the scenario 2.

Finally, we consider also the possibility that the state
(\ref{H1}) is the isotriplet $\Theta_{\bar c
1}^{(\mathbf{15})}(\frac32)$ (scenario 3).

We present in Table III predictions for the masses of all other
charmed pentaquarks following from each of the three assignments
described above. We present our predictions in terms of
spin-averaged masses for heavy quark doublets, defined as
\begin{eqnarray}\label{spav}
\langle \Theta_{\bar c 1}^{(R)} \rangle_{\rm sp-av}
= \frac13 \Theta_{\bar c 1}^{(R)}(\frac12) +
\frac23 \Theta_{\bar c 1}^{(R)}(\frac32) \,,\qquad
\langle \Theta_{\bar c 2}^{(R)} \rangle_{\rm sp-av}
\equiv \frac25 \Theta_{\bar c 2}^{(R)}(\frac32) +
\frac35 \Theta_{\bar c 2}^{(R)}(\frac52) \,.
\end{eqnarray}

\begin{center}
\begin{table}[ht!]
\begin{tabular}{|c|c|c|c|}
\hline
state & Scenario 1 & Scenario 2 & Scenario 3 \\
\hline
\hline
 & & & \\
$T_{\bar c 0}^{(\mathbf{3})}(\frac12)$ & $2896\pm 13\pm \delta_{1/m_c} $
& $2643\pm 25\pm \delta_{1/m_c}$   & $2744 \pm 20 \pm \delta_{1/m_c}$ \\
 & & & \\
$\langle T_{\bar c 1}^{(\mathbf{3})}\rangle$ & $2947$ & $2694$ & $2795$ \\
 & $====$ & $----$ & $----$  \\
$\langle \Theta_{\bar c 1}^{(\mathbf{6})}\rangle$ & $3099$ & $2846$ & $2947$ \\
 & $*--*$ & & ==== \\
$\Theta_{\bar c 0}^{(\mathbf{15})}(\frac12)$ & $3200$ & $2947$ & $3048$ \\
 & *==* & ==== & \\
$\langle \Theta_{\bar c 1}^{(\mathbf{15})}\rangle$ & $3251$ & $2998$ & $3099$ \\
 & & & $*--*$ \\
$\langle \Theta_{\bar c 2}^{(\mathbf{15})}\rangle$ & $3352$ & $3099$ & $3200$ \\
 & & *==* & *==* \\
$\langle \Theta_{\bar c 1}^{(\mathbf{15'})}\rangle$ & $3555$ & $3302$ & $3403$ \\
 & & & \\
\hline
\end{tabular}
{\caption{Predictions for the mass spectrum of the charmed
pentaquarks (in MeV) using the quark model with spin-color
hyperfine interaction. The three scenarios correspond to the
possible assignments of the 3099 MeV state seen by the H1 as
described in the text. The thresholds for strong two-body decays
are shown as $[D N]: ---- $, $[D^* N]: ==== $, $[D \Delta]: *--*
$, $[D^* \Delta]: *==* $. }}
\end{table}
\end{center}

We turn now to a discussion of these predictions. First, we
comment on the theoretical uncertainty in these estimates, which
were added in Table III only for the triplet states. These errors
include the experimental error in the H1 mass measurement
(\ref{H1}) and the error in the strength of the hyperfine
splitting $V=19\pm 1$ MeV. To this one should add also the
uncertainty $\delta_{1/m_c} \sim \Lambda^2/m_c \simeq 180$ MeV
coming from the hyperfine interaction with the heavy quark spin in
the state (\ref{H1}).

The lowest-lying states $T_{\bar c 0}^{(\mathbf{3})}(\frac12)$ in
the $\mathbf{3}$ of SU(3) are stable in all three scenarios
against strong decays into $[\bar D_s p]$ and $[\bar D \Lambda]$
(with thresholds at 2907 and 2985 MeV, respectively). Their masses
are somewhat higher than the previous estimate in Ref.~\cite{StWi}
of 2580 MeV. We note that the predictions in Table III assume
SU(3) symmetry. As a rough estimate of the SU(3) breaking, one
could add to these numbers an additional $\Delta_s = m_{\Xi_c} -
m_{\Lambda_c} = 180$ MeV. If this is done, the $T_{\bar c
0}^{(\mathbf{3})}(\frac12)$ states remain stable against strong
decays in scenario 2, but rise above the threshold for $[\bar D_s
p]$ in the other two scenarios. A search for a signal in this mass
region could help distinguish between the three scenarios.

Going to the higher mass states, in scenario 1 all
nonstrange states in the $\mathbf{\overline{6}}$ and $\mathbf{15}$
can decay strongly into $[\bar D^{(*)} p]_S$, and the $\mathbf{15'}$
decay into $[\bar D^{(*)} \Delta]_S$, with
decay widths of a typical size for a $S-$wave channel.
The scenario 3 is very similar.

The scenario 2 is more interesting. It contains three nonstrange states
stable under strong decays into $[\bar D^* p]_S$, which can only decay into
$[\bar D p]_S$. These are the
$\Theta_{\bar c 1}^{(\mathbf{\overline{6}})}(\frac12,\frac32)$ and
the $\Theta_{\bar c 0}^{(\mathbf{15})}(\frac12)$.
The heavy quark symmetry relations in Eq.~(\ref{Swave}) imply that two
of them should be narrow:
the $\Theta_{\bar c 1}^{(\mathbf{\overline{6}})}(\frac32)$ whose width
is suppressed by $\Lambda^2/m_c^2$, and the
$\Theta_{\bar c 0}^{(\mathbf{15})}(\frac12)$ whose width is reduced by a
factor of 4 due to the absence of the $\bar D^* p$ channel.
Thus, together with the
narrow $\Theta_{\bar c 2}^{(\mathbf{15})}(\frac32)$ identified with the H1 state
(\ref{H1}), this scenario contains 2 other narrow states well separated
at 2850 MeV ($\Theta_{\bar c 1}^{(\mathbf{\overline{6}})}(\frac32)$),
and at 2950 MeV ($\Theta_{\bar c 0}^{(\mathbf{15})}(\frac12)$).
This is a
distinctive experimental signature which should help distinguish this
assignment of the H1 state from
the other two proposed scenarios.

\section{Conclusions}

If exotic baryon states exist in nature, they add a new layer of
complexity to the hadronic spectrum, with a rich phenomenology.
The properties of these states can be studied in a model independent way
using the large $N_c$ expansion.
We extend the recent analysis of the positive parity exotics performed
in Ref.~\cite{JM3} to the negative parity states. In the quark model
these states correspond to the ground state with all quarks in $s$-wave
orbitals, and are thus expected to be lighter that their positive parity
counterparts.

Their spin-flavor structure is more complicated, corresponding to
a wave function transforming in the mixed symmetry representation.
In this respect, these states are closely related to orbitally
excited baryons, well studied in the large $N_c$ expansion
\cite{Goity,PY1,PY2,SU3}, and we make use of techniques developed
to deal with these states. We derive properties of the mass
spectrum of the exotic states in an expansion in $1/N_c$. Similar
methods can be applied to study other properties of the states,
such as their magnetic moments and strong decays.

We study both heavy (quark content $q^4 \bar Q$) and light $q^4\bar q$
pentaquarks, constructing the complete set of states for both SU(2)
and SU(3) flavor symmetry. In the heavy sector there are more states
than for the positive parity case, transforming as $\mathbf{3},
\mathbf{\overline{6}},\mathbf{15}$ and $\mathbf{15'}$ under SU(3).
We derive two model-independent mass relations Eqs.~(\ref{mr1}), (\ref{mr2})
to $O(1/N_c^2)$ connecting the masses of these states.

The mass spectrum of the negative parity light pentaquarks is richer, and
includes both exotic ($\mathbf{\overline{10},27,35}$) and non-exotic
SU(3) representations $(\mathbf{1,8,10}$). Even though mixing with the
orbitally excited regular baryons will likely affect the mass spectrum of
the nonexotic states, we study in some detail the mass spectrum of the
$\mathbf{1}$ states, which are expected to be the lightest pentaquarks.
We show that their mass spectrum and mixing are constrained in a very
specific way at leading order in $1/N_c$.

In contrast to the symmetric spin-flavor (positive parity) exotic
states studied in Ref.~\cite{JM3}, for this case the large $N_c$
expansion is somewhat less predictive, and does not connect their
properties to those of the ground state baryons. More predictive
power is obtained in the constituent quark model with color-spin
interactions, which is used to compute the mass spectrum of heavy
exotic states. We show that this approach is equivalent to the
large $N_c$ expansion, with a particular relation among the
coefficients of the $O(1/N_c)$ operators. Using the recent
observation by H1 of an anticharmed pentaquark state, we make
predictions for all other charmed pentaquarks, and point out
experimental signatures for the remaining states. We suggest one
natural explanation for the narrow width of the H1 state. Assuming
that it is identified with the $\mathbf{15}_\frac32$ with the spin
of the light degrees of freedom $s_\ell^{\pi_\ell} = 2^-$, its
strong decay width to $\bar D^{(*)} N$ is suppressed by
$\Lambda^2/m_c^2$ from heavy quark symmetry.
We present heavy quark symmetry predictions for the strong decay
width ratios in $\bar D^{(*)} N$ and $\bar D^{(*)} \Delta$, which
should be useful in constraining the quantum numbers of these
states.


\begin{acknowledgements}
We are grateful to Bob Jaffe, Elizabeth Jenkins and Aneesh Manohar
for very useful and stimulating discussions. While this paper was
being completed, we received Ref.~\cite{MW} which presents a
similar treatment of the heavy pentaquarks in the $1/N_c$
expansion. We thank Margaret Wessling for informing us of her work
before publication. The work of D. P. has been supported by the
U.S. Department of Energy (DOE) under the Grant No.
DOE-FG03-97ER40546 and by the NSF under grant PHY-9970781. The
work of C.S. was partially supported by ANPCYT (Argentina) grant
PICT03-08580 and by Fundaci\'on Antorchas. We thank the organizers
of the `Large $N_c$ QCD' program in ECT$^\star$ Trento, Italy,
where this work has been completed, for an enjoyable workshop.
\end{acknowledgements}

\newpage
\appendix

\section{Matrix elements on exotic states}

We list here the matrix elements of the $O(1)$ operators $G_{\bar q}^{ia} G_{q}^{ia}$
and $T^2$ appearing in the Hamiltonian of the light negative parity exotic states.

\begin{eqnarray}\nonumber
\begin{array}{c|c|c}
\hline \mbox{state} & \frac{1}{N_c} G_{\bar q}^{ia} G_{q}^{ia}  &
\frac{1}{N_c} T^a T^a\\
\hline
\hline
 & \\
\mathbf{1}_\frac12(\mathbf{3}_0) & 0 & \frac{N_c^2-9}{12N_c}\\
 &  \\
\mathbf{1}_\frac12(\mathbf{3}_1) & -\frac{3N_c+11}{24N_c} & \frac{N_c^2-9}{12N_c} \\
 & \\
\mathbf{1}_\frac12(\mathbf{3}_0)-\mathbf{1}_\frac12(\mathbf{3}_1)
&
\pm \frac{\sqrt3(N_c+5)}{16 N_c} & 0 \\
 & \\
\hline
 & \\
\mathbf{\overline{10}}_\frac12(\mathbf{\overline{6}}_1) &
\frac{1}{6 N_c} &
\frac{(N_c+3)(N_c+9)}{12N_c}\\
 & \\
\mathbf{\overline{10}}_\frac32(\mathbf{\overline{6}}_1) & -\frac{1}{12 N_c} & \frac{(N_c+3)(N_c+9)}{12N_c} \\
 &  \\
\hline
 &  \\
\mathbf{27}_\frac12(\mathbf{15}_0) & 0 & \frac{N_c^2+12N_c+51}{12N_c}\\
 &  \\
\mathbf{27}_\frac12(\mathbf{15}_1) & \frac{1}{6N_c} & \frac{N_c^2+12N_c+51}{12N_c}\\
 &  \\
\mathbf{27}_\frac12(\mathbf{15}_0)-\mathbf{27}_\frac12(\mathbf{15}_1) & 0 & 0 \\
 &  \\
\hline
 &  \\
\mathbf{27}_\frac32(\mathbf{15}_1) & -\frac{1}{12N_c} & \frac{N_c^2+12N_c+51}{12N_c}\\
 &  \\
\mathbf{27}_\frac32(\mathbf{15}_2) & \frac{1}{4N_c} & \frac{N_c^2+12N_c+51}{12N_c}\\
 &  \\
\mathbf{27}_\frac32(\mathbf{15}_1)-\mathbf{27}_\frac32(\mathbf{15}_2) & 0 & 0 \\
 &  \\
\hline
 & \\
\mathbf{27}_\frac52(\mathbf{15}_2) & -\frac{1}{6 N_c} & \frac{N_c^2+12 N_c+51}{12N_c}\\
 &  \\
\mathbf{35}_\frac12(\mathbf{15}'_1) & \frac{1}{6 N_c} & \frac{N_c^2+12 N_c+99}{12N_c} \\
 & \\
\mathbf{35}_\frac32(\mathbf{15}'_1) & -\frac{1}{12 N_c} & \frac{N_c^2+12 N_c+99}{12N_c} \\
 & \\
\hline
\end{array}
\end{eqnarray}

\end{document}